\documentclass[12pt]{iopart}
\usepackage{graphicx}

\begin{document}

\title{Theory of the radiation pressure on magneto--dielectric materials}

\author{Stephen M. Barnett}

\address{School of Physics and Astronomy, University of Glasgow, Glasgow G12 8QQ, UK}

\ead{stephen.barnett@glasgow.ac.uk} 

\author{Rodney Loudon}

\address{Computer Science and Electronic Engineering, University of Essex, Colchester CO4 3SQ, UK}

\ead{loudr@essex.ac.uk}

\date{\today}

\begin{abstract}
We present a classical linear response theory for a magneto--dielectric material and determine the polariton
dispersion relations.  The electromagnetic field fluctuation spectra are obtained and polariton sum rules for their 
optical parameters are presented.  The electromagnetic field for systems with multiple polariton branches is
quantised in 3 dimensions and field operators are converted to 1--dimensional forms appropriate for parallel
light beams.  We show that the field--operator commutation relations agree with previous calculations that ignored
polariton effects.  The Abraham (kinetic) and Minkowski (canonical) momentum operators are introduced and 
their corresponding single--photon momenta are identified.  The commutation relations of these and of their angular
analogues support the identification, in particular, of the Minkowski momentum with the canonical momentum of
the light.  We exploit the Heaviside--Larmor symmetry of Maxwell's equations to obtain, very directly, the 
Einsetin-Laub force density for action on a magneto--dielectric.  The surface and bulk contributions to the radiation
pressure are calculated for the passage of an optical pulse into a semi--infinite sample.
\end{abstract}

\maketitle

\section{Introduction}

At the heart of the problem of radiation pressure is the famous Abraham--Minkowski dilemma concerning
the correct form of the electromagnetic momentum in a material medium \cite{Minkowski1908,Abraham1909,
Abraham1910,AbrahamBecker1932}; a problem which, despite of its 
longevity, continues to attract attention \cite{BaxterLoudon2010,MilonniBoyd2010,BarnettLoudon2010,
Mansuripur2010,Kemp2011}.  The resolution of this dilemma lies is the identification of the two momenta,
due to Abraham and Minkowski, with the kinetic and canonical momenta of the light, respectively \cite{Barnett2010}.  
It serves to indicate, moreover, why the different momenta are apparent in different physical situations
\cite{Kemp2011}.  In this paper we shall be concerned with manifestations of optical momentum in radiation
pressure on media and, in particular, on magneto--dielectric media. 

Most existing work on the theory of radiation pressure, as reviewed in \cite{BaxterLoudon2010,MilonniBoyd2010,
BarnettLoudon2010,Mansuripur2010,Kemp2011}, treated non--magnetic materials  but there has been recent 
progress in the partial determination of the effects of material magnetisation.  The rise of interest in metamaterials 
and in particular those with negative refractive index adds urgency to addressing this point \cite{Milonnibook}.

We consider magneto--dielectric materials in which both the electric permittivity $\varepsilon$ and the magnetic 
permeability $\mu$ are isotropic functions of the angular frequency $\omega$.  The quantum theory of the 
electromagnetic field for such media is now well--developed, with analyses in the literature based on 
Green's functions \cite{Matloob2004,Judge2013} and also on Hopfield--like models based on coupling to
a harmonic polarisation and magnetisation field \cite{Kheirandish2006,Kheirandish2008,Kheirandish2009,Amooshahi2009}.  
The approach we shall adopt is to work with the elementary excitations within the medium, which are polarition coupled--modes 
of the electromagnetic field with the electric and magnetic resonances.  We develop the associated classical linear--response
theory and use this to determine the electric and magnetic field--fluctuation spectra.  The corresponding quantised field 
operators are expressed in terms of the polariton creation and destruction operators and we show that these satisfy
the required standard commutation relations.  We use these to construct the Minkowski and Abraham momentum operators
and also the corresponding angular momenta, and show how the commutators of these with the vector potential facilitate 
the interpretation of the rival momenta \cite{Barnett2010}.  The known form of the magnetic Lorentz force is confirmed and
used to explore the momentum transfer from light to a half--space sample and so provide the extension to permeable
medium of earlier work on dielectrics \cite{LBB2005}.


\section{Classical theory}

It suffices for our purposes to ignore the effects of absorption and so work with a real permittivity and permeability.
These arise naturally from the dispersion relations in a Hopfield--type model in which the electromagnetic field
is coupled to harmonic polarization and magnetisation fields associated with the host medium \cite{Kittel}.

\subsection{Linear response}

The relative permittivity and permeability of a (lossless) magneto--dielectric at angular frequency $\omega$ may be written
in the simple forms \cite{Hayes}
\begin{eqnarray}
\varepsilon(\omega) &=& \prod_{e=1}^{n_e} \frac{\omega^2_{{\rm L}e}-\omega^2}{\omega^2_{{\rm T}e}-\omega^2}  \nonumber \\
\mu(\omega) &=& \prod_{m=1}^{n_m}  \frac{\omega^2_{{\rm L}m}-\omega^2}{\omega^2_{{\rm T}m}-\omega^2} \, ,
\end{eqnarray}
where $n_e$ and $n_m$ are the numbers of electric and magnetic dipole resonances in the medium, associated
with longitudinal and transverse frequencies indicated by the subscripts L and T.  It follows that the values of the permittivity
and permeability at zero and infinite frequencies are
\begin{eqnarray}
\varepsilon(0) &=& \prod_{e=1}^{n_e} \frac{\omega^2_{{\rm L}e}}{\omega^2_{{\rm T}e}}\, ,  \qquad
\varepsilon(\infty) = 1 \nonumber \\
\mu(0) &=&  \prod_{m=1}^{n_m}  \frac{\omega^2_{{\rm L}m}}{\omega^2_{{\rm T}m}}\, ,
\qquad \mu(\infty) = 1 \, ,
\end{eqnarray}
in accord with the generalized Lyddane--Sachs--Teller relation \cite{CochranCowley1962}.  The phase refractive 
index is
\begin{equation}
\eta_{\rm p}(\omega) = \sqrt{\varepsilon(\omega)\mu(\omega)} \, ,
\end{equation}
as usual \cite{Jackson1999}.

We consider a magneto--dielectric medium in the absence of free charges and currents, so that our electric and
magnetic fields are governed by Maxwell's equations in the form 
\begin{eqnarray}
\label{Eq1.1}
\label{MaxwellEq}
\mbox{\boldmath$\nabla$}\cdot{\bf D} &=& 0  \nonumber \\
\mbox{\boldmath$\nabla$}\cdot{\bf B} &=& 0  \nonumber \\
\mbox{\boldmath$\nabla$}\times{\bf E} &=& -\frac{\partial {\bf B}}{\partial t} \nonumber \\
\mbox{\boldmath$\nabla$}\times{\bf H} &=& \frac{\partial {\bf D}}{\partial t} \, .
\end{eqnarray}
The electromagnetic fields described by these propagate as transverse plane--waves, with an evolution described by the complex 
factor $\exp({i\bf k}\cdot{\bf r} - i\omega t)$, with $k = |{\bf k}|$ and $\omega$ related by the appropriate dispersion relation.
The four complex electric and magnetic fields are related by \cite{Jackson1999}
\begin{eqnarray}
\label{PolMagRelns}
{\bf D}(\omega) &= \varepsilon_0{\bf E}(\omega) + {\bf P}(\omega) = \varepsilon_0 \varepsilon(\omega){\bf E}(\omega) \nonumber \\
{\bf B}(\omega) &= \mu_0[{\bf H}(\omega) + {\bf M}(\omega)] = \mu_0\mu(\omega){\bf H}(\omega) \, ,
\end{eqnarray}
where ${\bf P}$ and ${\bf M}$ are, respectively, the medium's polarization and magnetization.  Henceforth, for the sake of brevity, we omit explicit reference to the frequency from our fields, permittivities and permeabilities.  

Now suppose that external stimuli
${\bf p}$ and ${\bf m}$ with the same frequency are applied parallel to ${\bf E}$ and ${\bf H}$ respectively, so that the relations 
(\ref{PolMagRelns}) become
\begin{eqnarray}
\label{FieldResponse}
{\bf D} &=& \varepsilon_0\varepsilon{\bf E} + {\bf p} \nonumber \\
{\bf B} &=& \mu_0\mu{\bf H} + \mu_0{\bf m} \, .
\end{eqnarray}
It is convenient to introduce 6--component field and stimulus variables defined to be
\begin{eqnarray}
{\bf f} &=& ({\bf E}, Z_0{\bf H})  \nonumber \\
{\bf s} &=& V({\bf p}, {\bf m}/c) \, 
\end{eqnarray}
where $Z_0 = \sqrt{\mu_0/\varepsilon_0}$ is the usual free-space impedance and $V$ is a suitably chosen sample volume.
The energy of interaction between the field and the stimulus components is then
\begin{equation}
I = -\sum_{j=1}^6 f_is_i = -V({\bf E}\cdot{\bf p} + \mu_0{\bf H}\cdot{\bf m}) \, .
\end{equation}
The solutions for the field components obtained by substitution of (\ref{FieldResponse}) into Maxwell's equations can be written 
\begin{equation}
f_i = \sum_{j=1}^6 T_{ij}s_j
\end{equation}
where the linear response matrix is
\begin{equation}
\label{Tmatrix}
{\bf T} = \frac{1}{\varepsilon_0V{\rm Den}}
\left(\begin{array} {cccccc}
                              \omega^2\mu  & 0  &  0 & 0 & \omega ck_z & -\omega ck_y \\
                              0  & \omega^2\mu  & 0 & -\omega c k_z & 0 & \omega ck_x \\
                              0  &  0  &  \omega^2\mu & \omega ck_y & -\omega ck_x & 0 \\
                              0  &  -\omega ck_z & \omega ck_y & \omega^2\varepsilon & 0 & 0\\
                              \omega ck_z & 0 & -\omega ck_x & 0 & \omega^2\varepsilon & 0 \\
                             -\omega c k_y & \omega ck_x & 0 & 0 & 0 & \omega^2\varepsilon \end{array}\right) \, ,
\end{equation}
with the common denominator
\begin{equation}
\label{PolDen}
{\rm Den} = c^2k^2 - \varepsilon\mu \omega^2 \, ,
\end{equation}                              
the zeroes of which give the dispersion relation for the field.  The elements of the matrix ${\bf T}$ agree with and extend partial 
results obtained previously for non-magnetic media \cite{BarkerLoudon1972,LifshitzPitaevskii1980}.

\subsection{Field fluctuations}
\label{Sect2.2}

The frequency and wave--vector fluctuation spectra at zero temperature are obtained from the Nyquist formula 
\cite{BarkerLoudon1972}
\begin{equation}
\langle f_if_i \rangle_{\omega, {\bf k}} = \frac{\hbar}{\pi}{\rm Im}\left(T_{ij}\right) \, ,
\end{equation}
where the required imaginary parts occur automatically in magneto--dielectric models that include damping mechanisms
or, in the limit of zero damping, by the inclusion of a vanishingly small imaginary part in $\omega$.  In the latter case,
which applies to our model, we find
\begin{equation}
\label{ImDen}
{\rm Im}\left(\frac{1}{\rm Den}\right) \rightarrow \frac{\pi}{2kc^2}\left[\delta\left(k - \frac{\omega\eta_{\rm p}}{c}\right)
- \delta\left(k + \frac{\omega\eta_{\rm p}}{c}\right)\right] \, .
\end{equation}
The total fluctuations are obtained by integration
\begin{equation}
\label{Fluct3D}
\langle f_if_i \rangle = \int_0^\infty d\omega \langle f_if_i \rangle_\omega
= \int_0^\infty d\omega \frac{V}{(2\pi)^3} \int d{\bf k} \langle f_if_i \rangle_{\omega, {\bf k}}  \, ,
\end{equation}
where the frequency $\omega$ and the three--diemnsional (3D) wave vector ${\bf k}$ are taken as continuous and
independent variables.

Consider plane--wave propagation parallel to the $z$-axis in a sample of length $L$ and cross--sectional area $A$.
The frequency fluctuation spectra are then obtained from the one--dimensional (1D) version of (\ref{Fluct3D}) as
\begin{equation}
\langle f_if_i \rangle_\omega = \frac{\hbar L}{2\pi^2}\int_0^\infty dk \: {\rm Im}\left(T_{ij}\right) \, .
\end{equation}
The contributions of the component $x$-- and $y$--polarised waves give
\begin{eqnarray}
\left\langle E^2_x \right\rangle &= \left\langle E^2_y \right\rangle = \frac{\hbar\omega}{4\pi A}
\left(\frac{\mu_0\mu}{\varepsilon_0\varepsilon}\right)^{1/2} 
\nonumber \\
\left\langle H^2_y \right\rangle &= \left\langle H^2_x \right\rangle = \frac{\hbar\omega}{4\pi A}
\left(\frac{\varepsilon_0\varepsilon}{\mu_0\mu}\right)^{1/2} \, ,
\end{eqnarray}
with the use of (\ref{Tmatrix}) and (\ref{ImDen}).  The $E$ and $H$ spectra, of course, have the usual relative magnitude for a
magneto--dielectric material.  The total fluctuations are obtained by integration of the spectra over $\omega$ as in (\ref{Fluct3D}).

\subsection{Polariton modes}

The combined excitation modes of the material dipoles and the electromagnetic field are the polaritons \cite{MillsBurstein1974}.
The transverse polaritons for a material of cubic symmetry, as assumed here, are twofold degenerate corresponding to the two 
independent polarisation directions.  Their dispersion relation is determined by the poles in the linear response function, the 
components of which are given in (\ref{Tmatrix}).  The vanishing denominator (Den) gives
\begin{equation}
\label{Dispersion}
c^2k^2 = \varepsilon\mu\omega^2 = \eta_{\rm p}^2\omega^2 \, ,
\end{equation}
which is the desired transverse dispersion relation.  For our relative permittivity and permeability, there are $n_e + n_m + 1$
transverse polariton frequencies for each wave vector and these are independent of the direction of ${\bf k}$.  They jointly 
involve all of the resonant frequencies in the electric permittivity and the magnetic permeability.  It is convenient to enumerate 
the transverse polariton branches by a discrete index $u$, and there is no overlap in frequency $\omega_{{\bf k}u}$ between
the different branches.  There are also $n_e + n_m$ longitudinal modes at frequencies $\omega_{{\rm L}_e}$ and $\omega_{{\rm L}_m}$,
independent of the wave vector, but these will concern us no further.

Subsequent calculations involve the polariton group index $\eta_{\rm g}$, defined in terms of the phase index $\eta_{\rm p}$:
\begin{eqnarray}
\eta_{\rm g} = c\frac{dk}{d\omega} = d\frac{(\omega\eta_{\rm p})}{d\omega}
\end{eqnarray}
for $\eta_{\rm p} = ck/\omega$.  The two refractive indices satisfy a satisfy a variety of sum rules over the polariton branches,
given by \cite{Huttner1991,HuttnerBarnett1992,Barnett1995}
\begin{eqnarray}
\label{firstsumrules}
\sum_u \left(\frac{\eta_{\rm p}}{\eta_{\rm g}}\right)_{{\bf k}u} &=& 1 \nonumber \\
\sum_u \left(\frac{1}{\eta_{\rm p}\eta_{\rm g}}\right)_{{\bf k}u} &=& 1 
\end{eqnarray}
and by \cite{BarnettLoudon2012}
\begin{eqnarray}
\label{secondsumrules}
\sum_u \left(\frac{\varepsilon}{\eta_{\rm p}\eta_{\rm g}}\right)_{{\bf k}u} = 1 \nonumber \\
\sum_u \left(\frac{\mu}{\eta_{\rm p}\eta_{\rm g}}\right)_{{\bf k}u} = 1 \, ,
\end{eqnarray}
where all of the optical variables are evaluated at frequency $\omega_{{\bf k}u}$, so the sums run over every frequency $\omega_{{\bf k}u}$
that is a solution of the dispersion relation (\ref{Dispersion}) for a given wave vector ${\bf k}$.


\section{Quantum theory}

\subsection{Field quantisation}

The polaritons are bosonic modes, quantised by standard methods in terms of creation and destruction operators 
with the commutation relation 
\begin{equation}
\label{Eq3.1}
\left[\hat{a}_{{\bf k}u},\hat{a}^\dagger_{{\bf k}'u'}\right] = \delta({\bf k}-{\bf k}')\delta_{u,u'} \, ,
\end{equation}
The Dirac and Kronecker delta functions have their usual properties.  The electromagnetic field quantisation derived 
previously for dispersive dielectric media \cite{Milonni1995} 
can be adapted for our magneto--dielectric medium by insertion of polariton branch labels and by conversion to SI units 
and continuous wave vectors.  It is convenient to write all of the field operators in the form
\begin{equation}
\label{VecPot}
\hat{\bf A}({\bf r},t) = \hat{\bf A}^{(+)}({\bf r},t) + \hat{\bf A}^{(-)}({\bf r},t) \, ,
\end{equation}
where the second term is the Hermitian conjugate of the first.  The deduced form of the transverse (or Coulomb gauge)
vector potential operator is then given by
\begin{equation}
\label{VecPotPos}
\hat{\bf A}^{(+)}({\bf r},t) = \left(\frac{\hbar}{16\pi^3\varepsilon_0}\right)^{1/2}\sum_u \int d{\bf k}
\left(\frac{\mu}{\omega\eta_{\rm p}\eta_{\rm g}}\right)^{1/2}_{{\bf k}u} \hat{a}_{{\bf k}u}(t)
e^{i{\bf k}\cdot{\bf r}}{\bf e}_{{\bf k}}
\end{equation}
where
\begin{equation}
\hat{a}_{{\bf k}u}(t) = \hat{a}_{{\bf k}u}\exp(-i\omega_{{\bf k}u}t) \, .
\end{equation}
The unit polarisation vector ${\bf e}_{{\bf k}}$ is assumed to be the same for all of the polariton branches at wave vector
${\bf k}$.  The two transverse polarisations for each wave vector are not shown explicitly here but they are important for
the simplification of mode summations \cite{CraigThiru1984} and they need to be kept in mind.  For non-magnetic materials,
with $\mu = 1$, the vector potential (\ref{VecPotPos}) agrees with equation (5) from \cite{ABE1996} and also with equation
(A12) in \cite{BLPS1990} when the summation over polariton branches is removed.

The electric field, ${\bf E}$, and magnetic flux density or induction, ${\bf B}$, field operators are readily obtained from the vector 
potential as
\begin{eqnarray}
\label{EFieldOp}
\hat{\bf E}^{(+)}({\bf r},t) &=& -\frac{\partial}{\partial t}\hat{\bf A}^{(+)}({\bf r},t)  \nonumber \\
&=& i \left(\frac{\hbar}{16\pi^3\varepsilon_0}\right)^{1/2}\sum_u \int d{\bf k}
\left(\frac{\omega\mu}{\eta_{\rm p}\eta_{\rm g}}\right)^{1/2}_{{\bf k}u} \hat{a}_{{\bf k}u}(t)
e^{i{\bf k}\cdot{\bf r}}{\bf e}_{{\bf k}}
\end{eqnarray}
and
\begin{eqnarray}
\label{BFieldOp}
\hat{\bf B}^{(+)}({\bf r},t) &=& \mbox{\boldmath$\nabla$}\times\hat{\bf A}^{(+)}({\bf r},t)  \nonumber \\
&=& i \left(\frac{\hbar}{16\pi^3\varepsilon_0}\right)^{1/2}\sum_u \int d{\bf k}
\left(\frac{\mu}{\omega\eta_{\rm p}\eta_{\rm g}}\right)^{1/2}_{{\bf k}u} \hat{a}_{{\bf k}u}(t)
e^{i{\bf k}\cdot{\bf r}}{\bf k}\times{\bf e}_{{\bf k}} \, .
\end{eqnarray}
The remaining field operators, the displacement field ${\bf D}$ and the magnetic field ${\bf H}$, follow from the 
quantum equivalents of (\ref{PolMagRelns}) as 
\begin{equation}
\label{DFieldOp}
\hat{\bf D}^{(+)}({\bf r},t) = i \left(\frac{\hbar}{16\pi^3\varepsilon_0}\right)^{1/2}\sum_u \int d{\bf k}
\left(\frac{\omega\varepsilon\eta_{\rm p}}{\eta_{\rm g}}\right)^{1/2}_{{\bf k}u} \hat{a}_{{\bf k}u}(t)
e^{i{\bf k}\cdot{\bf r}}{\bf e}_{{\bf k}}
\end{equation}
and 
\begin{equation}
\label{HFieldOp}
\hat{\bf H}^{(+)}({\bf r},t) = i \left(\frac{\hbar}{16\pi^3\varepsilon_0}\right)^{1/2}\sum_u \int d{\bf k}
\left(\frac{1}{\omega\mu\eta_{\rm p}\eta_{\rm g}}\right)^{1/2}_{{\bf k}u} \hat{a}_{{\bf k}u}(t)
e^{i{\bf k}\cdot{\bf r}}{\bf k}\times{\bf e}_{{\bf k}} \, .
\end{equation}
The above four operator expressions are consistent with previous work \cite{Milonni1995}.  The field quantisation
for a magneto--dielectric medium has also been performed in a quantum--mechanical linear--response approach
\cite{Matloob2004,Judge2013} that assumes arbitrary complex forms for the permittivity and permeability.  The
resulting expressions for the $E$ and $B$ operators contain the polariton denominator (\ref{PolDen}) in their 
transverse parts.  A theory along these lines has been further developed by a more microscopic approach in 
which the explicit forms of the electric and magnetic susceptibilities are derived \cite{Kheirandish2006,Kheirandish2008}.

\subsection{Field commutation relations}

The validity of the field operators derived here (or at least their self--consistency) is demonstrated by the confirmation
that they satisfy the required equal--time commutation relations \cite{CraigThiru1984,Power1964}.  Thus it follows from
forms of our operators, given above, that 
\begin{eqnarray}
\label{ADCommutator}
\left[\hat{A}_i({\bf r}), -\hat{D}_j({\bf r}')\right] &=& \frac{i\hbar}{16\pi^3}\sum_u\int d{\bf k}
\left(\frac{\eta_{\rm p}}{\eta_{\rm g}}\right)_{{\bf k}u}\left[e^{i{\bf k}\cdot({\bf r}-{\bf r}') }+ 
e^{i{\bf k}\cdot({\bf r}'-{\bf r}) }\right]e_{{\bf k}i}e_{{\bf k}j} \nonumber \\
&=& \frac{i\hbar}{8\pi^3}\int d{\bf k}e^{i{\bf k}\cdot({\bf r}-{\bf r}')}e_{{\bf k}i}e_{{\bf k}j} \nonumber \\
&=& i\hbar \delta_{ij}^\perp({\bf r}-{\bf r}') \, ,
\end{eqnarray}
with recognition that the two exponents in the first step give the same contributions and a crucial use of the first 
sum rule from equation (\ref{firstsumrules}).  The transverse delta function, $\delta_{ij}^\perp({\bf r}-{\bf r}')$, in the
final step \cite{CraigThiru1984,Power1964,MilonniVacuum} relies on the implicit summation of the contributions of the 
two transverse polarisations.  The common time $t$ is omitted from the field operators in the commutators here and
subsequently as it does not appear in the final results.  An alternative canonical commutator is that for the vector 
potential and the electric field:
\begin{eqnarray}
\label{AECommutator}
\left[\hat{A}_i({\bf r}), -\varepsilon_0\hat{E}_j({\bf r}')\right] &=& \frac{i\hbar}{8\pi^3}\sum_u\int d{\bf k}
\left(\frac{\mu}{\eta_{\rm p}\eta_{\rm g}}\right)_{{\bf k}u}e^{i{\bf k}\cdot({\bf r}-{\bf r}')}
e_{{\bf k}i}e_{{\bf k}j} \nonumber \\
&=& i\hbar \delta_{ij}^\perp({\bf r}-{\bf r}') \, ,
\end{eqnarray}
with the use of a sum rule from equation (\ref{secondsumrules}).  It follows from the operator form of the first relation
in equation (\ref{PolMagRelns}) that the vectors potential and the polarisation commute:
\begin{equation}
\left[\hat{A}_i({\bf r}), \hat{P}_j({\bf r}')\right] = 0 \, .
\end{equation}
This is most satisfactory as the two operators are associated with properties of different physical entities: the electromagnetic
field and the medium respectively.  We note that the commutator in equation (\ref{ADCommutator}) has been given previously
\cite{Barnett2010} for non--magnetic dielectric media.

The remaining non--vanishing field commutators are
\begin{eqnarray}
\left[\hat{E}_i({\bf r}), \hat{B}_j({\bf r}') \right] &=& \frac{\hbar}{8\pi^3\varepsilon_0}\sum_u\int d{\bf k}
\left(\frac{\mu}{\eta_{\rm p}\eta_{\rm g}}\right)_{{\bf k}u}e^{i{\bf r}\cdot({\bf r}-{\bf r}')} \epsilon_{ijh}k_h \nonumber \\
&=& i\frac{\hbar}{\varepsilon_0}\epsilon_{ijh}\nabla'_h\delta({\bf r}-{\bf r}') \nonumber \\
\left[\hat{E}_i({\bf r}), \hat{H}_j({\bf r}') \right] &=& \frac{\hbar}{8\pi^3\mu_0\varepsilon_0}\sum_u\int d{\bf k}
\left(\frac{1}{\eta_{\rm p}\eta_{\rm g}}\right)_{{\bf k}u}e^{i{\bf r}\cdot({\bf r}-{\bf r}')} \epsilon_{ijh}k_h \nonumber \\
&=& i\frac{\hbar}{\mu_0\varepsilon_0}\epsilon_{ijh}\nabla'_h\delta({\bf r}-{\bf r}') \nonumber \\
\left[\hat{D}_i({\bf r}), \hat{B}_j({\bf r}') \right] &=& \frac{\hbar}{8\pi^3}\sum_u\int d{\bf k}
\left(\frac{\eta_{\rm p}}{\eta_{\rm g}}\right)_{{\bf k}u}e^{i{\bf r}\cdot({\bf r}-{\bf r}')} \epsilon_{ijh}k_h \nonumber \\
&=& i\hbar\epsilon_{ijh}\nabla'_h\delta({\bf r}-{\bf r}') \nonumber \\
\left[\hat{D}_i({\bf r}), \hat{H}_j({\bf r}') \right] &=& \frac{\hbar}{8\pi^3\mu_0}\sum_u\int d{\bf k}
\left(\frac{\varepsilon}{\eta_{\rm p}\eta_{\rm g}}\right)_{{\bf k}u}e^{i{\bf r}\cdot({\bf r}-{\bf r}')} \epsilon_{ijh}k_h 
\nonumber \\
&=& i\frac{\hbar}{\mu_0}\epsilon_{ijh}\nabla'_h\delta({\bf r}-{\bf r}') \, ,
\end{eqnarray}
where we have used the sum rules given in equations (\ref{firstsumrules}) and (\ref{secondsumrules}).  Here 
$\epsilon_{ijh}$ is the familiar permutation symbol \cite{Stephenson1990,BarnettRadmore1997} and the repeated
index $h$ is summed over the three cartesian coordinates $x$, $y$ and $z$.  The first of these results generalises
a commutator previously derived for fields in vacuo \cite{Power1964}.  Note that together these commutation relations
require that the polarisation operator commutes with the magnetic field operator and that the magnetisation also
commutes with the electric field operator:
\begin{eqnarray}
\left[\hat{P}_i({\bf r}), \hat{H}_j({\bf r}') \right] &=& 0 \nonumber \\
\left[\hat{M}_i({\bf r}), \hat{E}_j({\bf r}') \right] &=& 0 \, ,
\end{eqnarray}
which is a physical consequence of the fact that the two pairs of operators in each commutator are associated with 
properties of different systems: the medium and the electromagnetic field.

\subsection{Parallel beams: 3D to 1D conversion}
\label{Sect3.3}

For parallel light beams, it is sometimes convenient to work with field operators defined for dependence on a single spatial
coordinate $z$ with a one--dimensional wave vector $k$.  Thus, for a beam of cross--sectional area $A$, conversion from 
3D to 1D is achieved by making the substitutions
\begin{eqnarray}
\int d{\bf k} &\rightarrow & \frac{4\pi^2}{A}\int dk   \nonumber \\
\delta({\bf k} - {\bf k}') &\rightarrow & \frac{A}{4\pi^2}\delta(k - k') \nonumber \\
\hat{a}_{{\bf k}u} &\rightarrow & \frac{\sqrt{A}}{2\pi} \: \hat{a}_{ku} \, .
\end{eqnarray}
The vector potential operator (\ref{VecPotPos}) for polarisation parallel to the $x$--axis is converted in this way to
\begin{equation}
\label{Eq3.15}
\hat{\bf A}^{(+)}(z,t) = \left(\frac{\hbar}{4\pi\varepsilon_0A}\right)^{1/2}\sum_u \int dk 
\left(\frac{\mu}{\omega\eta_{\rm p}\eta_{\rm g}}\right)^{1/2}_{ku} \hat{a}_{ku}(t)e^{ikz}\: \tilde{\bf x} \, ,
\end{equation}
where $\tilde{\bf x}$ is the unit vector in the $x$--direction.  The orthogonal degenerate polariton modes give a 
vector potential parallel to $\tilde{\bf y}$.  The four field operators retain their forms given in equations (\ref{EFieldOp})
to (\ref{HFieldOp}) except for appropriate changes in the first square--root factors and vector directions.  
The electric and
displacement fields are parallel to the $x$--axis, while the magnetic field and induction are parallel to the $y$--axis.
The commutation relation for the creation and destruction operators retains the form given in equation (\ref{Eq3.1})
but with ${\bf k}$ replaced by $k$.  For $\mu = 1$ and a single--resonance dielectric, the vector potential (\ref{Eq3.15})
agrees with equation (3.25) in \cite{Barnett1995} and equation (9) in \cite{Huttner1991}.

The single--coordinate vector potential can also be expressed as an integral over frequency by means of the 
conversions
\begin{eqnarray}
\int dk &\rightarrow & \int d\omega \frac{\eta_{\rm g}}{c} \nonumber \\
\delta(k-k') &\rightarrow & \frac{c}{\eta_{\rm g}}\delta(\omega - \omega') \nonumber \\
\hat{a}_{ku} &\rightarrow & \left(\frac{c}{\eta_{\rm g}}\right)^{1/2}\hat{a}_\omega \, .
\end{eqnarray}
There are no overlaps in frequency between the different twofold--degenerate polariton branches and the summation
over $u$ is accordingly removed from the vector potential, which becomes
\begin{equation}
\hat{\bf A}^{(+)}(z,t) = \int d\omega \left(\frac{\hbar}{4\pi\omega A}\right)^{1/2}
\left(\frac{\mu_0\mu}{\varepsilon_0\varepsilon}\right)^{1/4}\hat{a}_\omega(t) \exp\left(i\omega\eta_{\rm p}z/c\right)
\: \tilde{\bf x} \, ,
\end{equation}
where 
\begin{equation}
\hat{a}_\omega(t) = \hat{a}_\omega e^{-i\omega t} \, .
\end{equation}
This destruction operator and the associated creation operator satisfy the continuum commutation relation
\begin{equation}
\left[\hat{a}_\omega, \hat{a}^\dagger_{\omega'}\right] = \delta(\omega - \omega') \, .
\end{equation}
The field operators obtained by conversion in this way of (\ref{EFieldOp}) to (\ref{HFieldOp}) are
\begin{eqnarray}
\label{1DOps}
\hat{\bf E}^{(+)}(z,t) &=& i\int d\omega \left(\frac{\hbar\omega}{4\pi A}\right)^{1/2}
\left(\frac{\mu_0\mu}{\varepsilon_0\varepsilon}\right)^{1/4} \hat{a}_\omega(t) 
\exp\left(i\omega \eta_{\rm p}z/c\right) \tilde{\bf x} \nonumber \\
\hat{\bf B}^{(+)}(z,t) &=& i\int d\omega \left(\frac{\hbar\omega}{4\pi A}\right)^{1/2}
(\mu_0\mu)^{1/4}(\varepsilon_0\varepsilon)^{3/4} \hat{a}_\omega(t) 
\exp\left(i\omega \eta_{\rm p}z/c\right)  \tilde{\bf y} \nonumber \\
\hat{\bf D}^{(+)}(z,t) &=& i\int d\omega \left(\frac{\hbar\omega}{4\pi A}\right)^{1/2}
(\varepsilon_0\varepsilon)^{3/4}(\mu_0\mu)^{1/4} \hat{a}_\omega(t) 
\exp\left(i\omega \eta_{\rm p}z/c\right) \: \tilde{\bf x} \nonumber \\
\hat{\bf H}^{(+)}(z,t) &=& i\int d\omega \left(\frac{\hbar\omega}{4\pi A}\right)^{1/2}
\left(\frac{\varepsilon_0\varepsilon}{\mu_0\mu}\right)^{1/4} \hat{a}_\omega(t) 
\exp\left(i\omega \eta_{\rm p}z/c\right) \tilde{\bf y} \, .
\end{eqnarray}
There is again agreement with previously defined expressions \cite{Barnett1995,Huttner1991} for $\mu = 1$.
We can use these field operators to calculate the vacuum fluctuations and find
\begin{eqnarray}
\langle 0|\hat{E}^2_x(z,t)|0\rangle &=& 
\langle 0|\hat{E})_x^{(+)}(z,t)\hat{E})_x^{(-)}(z,t)|0\rangle = 
\int d\omega \frac{\hbar\omega}{4\pi A}\left(\frac{\mu_0\mu}{\varepsilon_0\varepsilon}\right)^{1/2} \nonumber \\
&=& \langle 0|\hat{E}^2_y(z,t)|0\rangle \nonumber \\
\langle 0|\hat{H}^2_y(z,t)|0\rangle &=& 
\langle 0|\hat{E})_y^{(+)}(z,t)\hat{H})_y^{(-)}(z,t)|0\rangle = 
\int d\omega \frac{\hbar\omega}{4\pi A}\left(\frac{\varepsilon_0\varepsilon}{\mu_0\mu}\right)^{1/2} \nonumber \\
&=& \langle 0|\hat{H}^2_x(z,t)|0\rangle \, ,
\end{eqnarray}
in exact agreement with results obtained in section \ref{Sect2.2}.

We shall make use of the 1D field operators derived here to calculate the force exerted by a photon on
a magneto--dielectric medium, but first return to the full 3D description to investigate the electromagnetic 
momentum.


\section{Momentum operators} 

\subsection{Minkowski and Abraham}

The much debated Abraham--Minkowski dilemma is most simply stated as a question: which of two
eminently plausible momentum densities, ${\bf D}\times{\bf B}$ and $c^{-2}{\bf E}\times{\bf H}$, is
the true or preferred value \cite{BarnettLoudon2010,Kemp2011}?  We amplify upon the answer 
given in \cite{Barnett2010}, making special reference to the effects of a magneto--dielectric medium.

Two rival forms for the electromagnetic energy--momentum tensors were derived by Minkowski \cite{Minkowski1908}
and Abraham \cite{Abraham1909,Abraham1910,AbrahamBecker1932}.  Their original formulations considered
electromagnetic fields in moving bodies, but it suffices for our our purposes to set the material velocity equal to 
zero.  These results continue to hold for the magneto--dielectric media of interest to us.  The two formulations differ principally
in their expressions for the electromagnetic momentum and we consider here the respective quantised versions
of these.

The Minkowski momentum is quite difficult to find in the paper \cite{Minkowski1908}, but it can be deduced trom his
expressions for other quantities.  Its quantised form, is represented by the operator
\begin{eqnarray}
\label{MinkMomOp}
\hat{\bf G}^{\rm M}(t) &=& \int d{\bf r} \hat{\bf D}({\bf r},t)\times\hat{\bf B}({\bf r},t) \nonumber \\
&=& \frac{\hbar}{2}\sum_{u,u'}\int d{\bf k} \left(\frac{\omega\varepsilon\eta_{\rm p}}{\eta_{\rm g}}\right)^{1/2}_{{\bf k}u}
\left(\frac{\mu}{\omega\eta_{\rm p}\eta_{\rm g}}\right)^{1/2}_{{\bf k}u'}
 \left[\hat{a}_{{\bf k}u}(t)\hat{a}^\dagger_{{\bf k}u'}(t) + \hat{a}^\dagger_{{\bf k}u}(t)\hat{a}_{{\bf k}u'}(t) \right. \nonumber\\
& & \qquad \left. + \hat{a}_{{\bf k}u}(t)\hat{a}_{-{\bf k}u'}(t) + \hat{a}^\dagger_{{\bf k}u}(t)\hat{a}^\dagger_{-{\bf k}u'}(t)\right]{\bf k} \, ,
\end{eqnarray}
where equations (\ref{BFieldOp}) and (\ref{DFieldOp}) have been used.  The diagonal part of this momentum operator
is
\begin{equation}
\hat{\bf G}^{\rm M}_{\rm diag} = \frac{1}{2}\sum_u \int d{\bf k} \left(\frac{\eta_{\rm p}}{\eta_{\rm g}}\right)_{{\bf k}u}
\left(\hat{a}^\dagger_{{\bf k}u}\hat{a}_{{\bf k}u} + \hat{a}_{{\bf k}u}\hat{a}^\dagger_{{\bf k}u}\right)\hbar{\bf k} \, .
\end{equation}
The fact that $k = \eta_{\rm p}\omega/c$ leads us to identify a single--photon momentum 
\begin{equation}
\label{Eq4.3}
\label{GarrChMom}
p_{({\rm M})} = \frac{\hbar\omega}{c}\left(\frac{\eta^2_{\rm p}}{\eta_{\rm g}}\right)_{{\bf k}u}
\end{equation}
with the individual polariton mode ${\bf k}u$.  This form of the Minkowski single--photon momentum has been
derived previously \cite{GarrisonChiao2004} for a single--resonance non--magnetic material.  There are good
reasons, however as we shall see below, \emph{not} to assign this value to the Minkowski momentum.

The corresponding form of the Abraham momentum operator, proportional to the Poynting vector for energy flow,
is \cite{Abraham1909,Abraham1910,AbrahamBecker1932}
\begin{eqnarray}
\label{AbMomOp}
\hat{\bf G}^{\rm A}(t) &=& \frac{1}{c^2}\int d{\bf r} \hat{\bf E}({\bf r},t)\times\hat{\bf H}({\bf r},t) \nonumber \\
&=& \frac{\hbar}{2}\sum_{u,u'}\int d{\bf k} \left(\frac{\omega\mu}{\eta_{\rm p}\eta_{\rm g}}\right)^{1/2}_{{\bf k}u}
\left(\frac{1}{\omega\mu\eta_{\rm p}\eta_{\rm g}}\right)^{1/2}_{{\bf k}u'}
 \left[\hat{a}_{{\bf k}u}(t)\hat{a}^\dagger_{{\bf k}u'}(t) + \hat{a}^\dagger_{{\bf k}u}(t)\hat{a}_{{\bf k}u'}(t) \right. \nonumber\\
& & \qquad \left. + \hat{a}_{{\bf k}u}(t)\hat{a}_{-{\bf k}u'}(t) + \hat{a}^\dagger_{{\bf k}u}(t)\hat{a}^\dagger_{-{\bf k}u'}(t)\right]{\bf k} \, ,
\end{eqnarray}
which differs from the corresponding expression for $\hat{\bf G}^{\rm M}(t)$ only in the the two square--root factors
that occur in the field operators (\ref{EFieldOp}) and (\ref{HFieldOp}). The diagonal part of this momentum operator
is
\begin{equation}
\hat{\bf G}^{\rm A}_{\rm diag} = \frac{1}{2}\sum_u \int d{\bf k} \left(\frac{1}{\eta_{\rm p}\eta_{\rm g}}\right)_{{\bf k}u}
\left(\hat{a}^\dagger_{{\bf k}u}\hat{a}_{{\bf k}u} + \hat{a}_{{\bf k}u}\hat{a}^\dagger_{{\bf k}u}\right)\hbar{\bf k} \, ,
\end{equation}
so that the associated single--photon momentum is
\begin{equation}
\label{pA}
p_{\rm A} = \frac{\hbar\omega}{c}\left(\frac{1}{\eta_{\rm g}}\right)_{{\bf k}u} \, ,
\end{equation}
which is the usual Abraham value.

The subscript M appears in brackets in equation (\ref{GarrChMom}) because there is an alternative form for
the Minkowski momentum, given by
\begin{equation}
\label{pM}
p_{\rm M} = \frac{\hbar\omega}{c}\eta_{\rm p} 
\end{equation}
is observed in experiments sufficiently accurate to distinguish between the phase and group refractive indices,
particularly the submerged mirror measurements of Jones and Leslie \cite{JonesLeslie1978}.  Note that the
difference between the two candidate momenta. $p_{({\rm M}}$ and $p_{\rm M}$ can be very large and even, the
case of media with a negative refractive index, point in opposite directions \cite{Scullion2008}.  The resolution
of the apparent conflict between the two forms of Minkowski momentum is discussed in section \ref{Sect4.3}.

\subsection{Vector potential--momentum commutators}

It is instructive to evaluate the commutators of the two momentum operators with the vector potential.  We 
can do this either by using the expressions for the fields in terms of the polariton creation and destruction 
operators or, more directly, by using the field commutation relations (\ref{ADCommutator}) and 
(\ref{AECommutator}), together with the fact that the vector potential commutes with both the magnetic
field and the induction.  

For the Minkowski momentum we find
\begin{eqnarray}
\label{MinkComm}
\left[\hat{A}_i({\bf r}), G^{\rm M}_j\right] &=& \epsilon_{jkl}\int d{\bf r}' \left[\hat{A}_i({\bf r}),\hat{D}_k({\bf r}')\right]
\hat{B}_l({\bf r}') \nonumber \\
&=& -i\hbar \epsilon_{jkl}\int d{\bf r}' \delta_{ik}^\perp({\bf r} - {\bf r}')\hat{B}_l({\bf r}') \nonumber \\
&=& -i\hbar \epsilon_{jkl}\epsilon_{lmn}\int d{\bf r}' \delta_{ik}^\perp({\bf r} - {\bf r}')\nabla'_m \hat{A}_n({\bf r}') \nonumber \\
&=& -i\hbar \nabla_j \hat{A}_i({\bf r}) + i\hbar \int d{\bf r}' \delta_{ik}^\perp({\bf r} - {\bf r}')\nabla'_k \hat{A}_j({\bf r}') \nonumber \\
&=& -i\hbar \nabla_j \hat{A}_i({\bf r}) \, ,
\end{eqnarray}
where we have used the summation convention so that repeated indices are summed over the three cartesian
directions.  It should
also be noted that the remaining fields will have the same form of commutation relation with the Minkowski
momentum, for example for the electric field we have
\begin{equation}
\left[\hat{E}_i({\bf r}), \hat{G}^{\rm M}_j\right] = -i\hbar \nabla_j \hat{E}_i({\bf r}) \, ,
\end{equation}
This follows directly from the relationships between these fields and the vector potential together with the fact that 
the vector potential commutes with the polarisation and the magnetisation.

For the Abraham momentum we can exploit our calculation for the Minkowski momentum to find the commutator
\begin{eqnarray}
\left[\hat{A}_i({\bf r}), \hat{G}^{\rm A}_j\right] &=& \epsilon_{jkl}\int d{\bf r}' \left[\hat{A}_i({\bf r}),\varepsilon_0\hat{E}_k({\bf r}')\right]
\mu_0\hat{H}_l({\bf r}') \nonumber \\
&=&  \epsilon_{jkl}\int d{\bf r}' \left[\hat{A}_i({\bf r}),\varepsilon_0\hat{E}_k({\bf r}')\right]
\left(\hat{B}_l({\bf r}') - \mu_0\hat{M}_l({\bf r}')\right) \nonumber \\
&=& -i\hbar \nabla_j \hat{A}_i({\bf r})
+i\hbar \epsilon_{jkl}\int d{\bf r}' \delta_{ik}^\perp({\bf r} - {\bf r}')\mu_0\hat{M}_l({\bf r}') \, .
\end{eqnarray} 
The second term, with its integration over the magnetisation, means that the commutator depends on both
a field property, the vector potential, and a medium property, the magnetisation \cite{Barnett2010}.

\subsection{Interpretation}
\label{Sect4.3}

The identification of the Minkowski and Abraham photon momenta respectively with the electromagnetic canonical
and kinetic momenta has been proposed in the past \cite{GarrisonChiao2004,BBL1993,Leonhardt2006,HindsBarnett2009},
but rigorously proven only more recently \cite{BarnettLoudon2010,Barnett2010}.  We need add here only a few brief remarks.

The commutation relation (\ref{MinkComm}) satisfied by the Minkowski momentum operator resembles the familiar
canonical commutator of the particle momentum operator,
\begin{equation}
\left[F_i({\bf r}),\hat{p}_j\right] = i\hbar \nabla_jF_i({\bf r}) \, ,
\end{equation}
where ${\bf F}({\bf r})$ is an arbitrary vector function of position.  Thus, analogously to its particle counterpart, the
Minkowski momentum operator for the electromagnetic field generates a spatial translation, in this case of the 
vector potential and the electric and magnetic fields.  The operator therefore indeed represents the canonical momentum
of the field and it is the observed momentum in experiments that measure the displacement of a body embedded in
a material host, as has been seen for a mirror immersed in a dielectric liquid \cite{JonesLeslie1978}, for the transfer of
momentum to charge carriers in the photon drag effect \cite{Gibson1980} and in the recoil of an atom in a host gas
\cite{Campbell2005}. The simple spatial derivative that occurs on the right of equation (\ref{MinkComm}) shows that the
measured single--photon momentum should have the Minkowski form in (\ref{pM}) and not that given in equation
(\ref{GarrChMom}).

The kinetic momentum of a material body is the simple product of its mass and velocity.  The form of the Abraham 
single--photon momentum in equation (\ref{pA}) is verified by thought experiments of the Einstein--box variety 
\cite{Balazs1953,Loudon2004}.  These use the principle of uniform motion of the centre of mass--energy as 
a single--photon pulse passes through a transparent dielectric slab and they reliably produce the Abraham
momentum.  The calculations remain valid with no essential modifications when the slab is made from a 
magneto--dielectric material.

More detailed analyses of the coupled material and electromagnetic momenta \cite{BarnettLoudon2010,Barnett2010}
show that the total momentum is unique but that this can be formed as the sum of alternative material and
electromagnetic field contributions
\begin{equation}
\hat{\mbox{\boldmath$\mathcal{P}$}}^{\rm canonical}_{\rm medium} + \hat{\bf G}^{\rm M} =
\hat{\mbox{\boldmath$\mathcal{P}$}}^{\rm kinetic}_{\rm medium} + \hat{\bf G}^{\rm A} \, ,
\end{equation}
where the $\hat{\mbox{\boldmath$\mathcal{P}$}}$ operators represent the collective momenta of all the electric and
magnetic dipoles that constitute the medium.  The total momentum is the same for both the canonical and kinetic 
varieties, both being conserved in the interactions between electromagnetic waves and material media.

\subsection{Angular momentum}

The electromagnetic field carries not only energy and linear momentum but also angular momentum and it is natural to
introduce angular momenta derived from the Minkowski and Abraham momenta in the forms
\begin{eqnarray}
\hat{\bf J}^{\rm M} &=& \int d{\bf r} \: {\bf r}\times\left[\hat{\bf D}({\bf r})\times\hat{\bf B}({\bf r})\right] \nonumber \\
\hat{\bf J}^{\rm A} &=& \frac{1}{c^2}\int d{\bf r} \: {\bf r}\times\left[\hat{\bf E}({\bf r})\times\hat{\bf H}({\bf r})\right] \, .
\end{eqnarray}
A careful analysis of a light beam carrying angular momentum entering a dielectric medium shows that,
in contrast with the linear momentum, the Minkowski angular momentum is the same inside and outside
the medium, but that the Abraham angular momentum is reduced in comparison to its free--space value
by the product of the phase and group indices \cite{PBL2003}.  The analogue of the Einstein--box argument
suggests that light carrying angular momentum entering a medium exerts a torque on it, inducing a 
rotation on propagation through it.  An object imbedded in the host, however, may be expected to experience
the influence of the same angular momentum as in free space and, indeed, this is what is seen in 
experiment \cite{KristensenWoerdman1994}.

The canonical or Minkowski angular momentum should be expected to induce a rotation of the electromagnetic
fields, which requires both a rotation of the coordinate and also of the direction of the field.  The requirement
to provide both of these transformations provides a stringent test of the identification of the Minkowski and
canonical momenta.  It is convenient to first rewrite the Minkowski linear momentum density in a new form:
\begin{eqnarray}
\left[\hat{\bf D}\times\hat{\bf B}\right]_i &=& \hat{D}_j\nabla_i\hat{A}_j - \hat{D}_j\nabla_j\hat{A}_i \nonumber \\
&=& \hat{D}_j\nabla_i\hat{A}_j - \nabla_j\left(\hat{D}_j\hat{A}_i\right) \, ,
\end{eqnarray}
where we have used the first Maxwell equation, $\nabla_j\hat{D}_j = 0$.  We can insert this form into our
expression for $\hat{\bf J}^{\rm M}$ and, on performing an integration by parts and discarding a physically--unimportant
boundary term we find
\begin{equation}
\hat{\bf J}^{\rm M} = \int d{\bf r}\left[\hat{D}_j ({\bf r}\times\mbox{\boldmath$\nabla$})\hat{A}_j + 
\hat{\bf D}\times\hat{\bf A}\right] \, , 
\end{equation}
which, in the absence of the medium reduces to the form obtained by Darwin \cite{Darwin1932,CTDRG1989}.

It is tempting, even natural, to associate the two contributions in the integrand with the orbital and spin 
angular momentum components of the total angular momentum.  This is indeed reasonable, but it should
be noted that neither part alone is a true angular momentum \cite{vanEnk1994a,vanEnk1994b,RotOAM}.  
It is instructive to consider the commutation relation with a single component of the angular momentum 
and so consider the operator $\mbox{\boldmath$\theta$}\cdot\hat{\bf J}^{\rm M}$:
\begin{eqnarray}
\label{AJComm}
\left[\hat{\rm A}(\bf r),\mbox{\boldmath$\theta$}\cdot\hat{\bf J}^{\rm M}\right] &=& 
-i\hbar\left[\mbox{\boldmath$\theta$}\cdot({\bf r}\times\mbox{\boldmath$\nabla$})\hat{\bf A}\right]^\perp
+ i\hbar(\mbox{\boldmath$\theta$}\times\hat{\bf A})^\perp \nonumber \\
&=& -i\hbar\left[\mbox{\boldmath$\theta$}\cdot({\bf r}\times\mbox{\boldmath$\nabla$})\hat{\bf A}
- \mbox{\boldmath$\theta$}\times\hat{\bf A}\right] \, .
\end{eqnarray}
The orbital and spin parts rotate, as far as is possible given the constraints of transversality, the amplitude
and direction of the potential \cite{vanEnk1994a,vanEnk1994b,RotOAM}.  
The combination of both of these gives the required transformation.  The 
commutator (\ref{AJComm}) gives the first order rotation of the vector potential about an axis parallel
to {\boldmath$\theta$} through the small angle $\theta$, as the canonical angular momentum should.


\section{Magnetic Lorentz force}

It remains to determine the radiation pressure  due to a light field  on our magneto--dielectric medium.
To complete this task we adopt the method used previously of evaluating the force exerted by a single--photon
plane--wave pulse normally incident on the medium \cite{LBB2005,Loudon2002}.  Before we can complete the 
calculation, however, we need to determine a suitable form for the electromagnetic force density.

\subsection{Heaviside--Larmor symmetry}

Maxwell's equations in the absence of free charges and currents (\ref{MaxwellEq})  exhibit the 
so--called Heaviside--Larmor symmetry \cite{Heaviside1892,Larmor1897}, with the forms that are 
invariant under rotational duality transformations given by \cite{Jackson1999}
\begin{eqnarray}
\label{HLTrans}
{\bf E} &=& {\bf E}'\cos\xi + Z_0{\bf H}'\sin\xi \nonumber \\
{\bf H} &=&  {\bf H}'\cos\xi-Z^{-1}_0{\bf E}' \sin\xi  \nonumber \\
{\bf D} &=& {\bf D}'\cos\xi + Z_0^{-1}{\bf B}'\sin\xi \nonumber \\
{\bf B} &=& {\bf B}'\cos\xi-Z_0{\bf D}' \sin\xi   \, ,
\end{eqnarray}
for any value of $\xi$ and where $Z_0$ is again the impedance of free space.  It is readily verified that
the four Maxwell equations are converted to the same set of equations in the primed fields.  The various
physical properties of the electromagnetic field must also be unchanged by the transformations 
\cite{Rose1955}.  We note, in particular, that the Minkowski and Abraham momentum operators, 
$\hat{\bf G}^{\rm M}$ and $\hat{\bf G}^{\rm A}$ given in equations (\ref{MinkMomOp}) and 
(\ref{AbMomOp}) and also the usual expressions for the electromagnetic field energy density
and Poynting vector are all invariant under the transformation (\ref{HLTrans}).

The standard form of the Lorentz force law in a non--magnetic dielectric is \cite{BarnettLoudon2006}
\begin{equation}
\label{LForce}
{\bf f}^{\rm L} = \left({\bf P}\cdot\mbox{\boldmath$\nabla$}\right){\bf E} + \mu_0\dot{\bf P}\times{\bf H} \, ,
\end{equation}
with terms proportional to the electric polarisation and its time derivative\footnote{This is usually written
in terms of the magnetic induction as \cite{BarnettLoudon2006}
\begin{equation*}
{\bf f}^{\rm L} = \left({\bf P}\cdot\mbox{\boldmath$\nabla$}\right){\bf E} + \dot{\bf P}\times{\bf B} \, .
\end{equation*}
In a magnetic medium, however, we need to distinguish between $\mu_0{\bf H}$ and ${\bf B}$.  That 
it should be the former that appears in the force density
follows on consideration of the screening effect of surrounding magnetic dipoles
in the medium, in much the same way as electric dipoles screen the electric field in a medium.}.  For a
magneto--dielectric medium we need to add the force due to the magnetisation and to do so in a 
manner that gives a force density that is invariant under the Heaviside--Larmor transformation.  It
follows from the transformation (\ref{HLTrans}) that the polarisation and magnetisation are similarly
transformed:
\begin{eqnarray}
{\bf P} &=& {\bf P}'\cos\xi + c^{-1}{\bf M}'\sin\xi \nonumber \\
{\bf M} &=& {\bf M}'\cos\xi-c{\bf P}' \sin\xi \, .
\end{eqnarray}
The required form of the force density, satisfying the Heaviside--Larmor symmetry is 
\begin{equation}
\label{ELForce}
{\bf f}^{\rm EL} = \left({\bf P}\cdot\mbox{\boldmath$\nabla$}\right){\bf E} + \mu_0\dot{\bf P}\times{\bf H}
+ \mu_0\left({\bf M}\cdot\mbox{\boldmath$\nabla$}\right){\bf H} - \varepsilon_0\mu_0\dot{\bf M}\times{\bf E} \, .
\end{equation}
The invariance of this expression under the Heavisde--Larmor transformation is easily shown.  This form
of the force density was derived by Einstein and Laub over 100 years ago \cite{EinsteinLaub1908} and there
have since been several independent re--derivations of it 
\cite{PenfieldHaus1967,Haus1969,Robinson1975,Mansuripur2007,Mansuripur2008,MansuripurZakharian2009}.
The final term in equation (\ref{ELForce}) has been given special attention in \cite{ShockleyJames1967},
where it is treated as a manifestation of the so--called `hidden momentum' given by 
$\varepsilon_0\mu_0{\bf M}\times{\bf E}$.  This line of thought has attracted a series of publications, with
several listed on page 616 of \cite{Jackson1999}.  Omission of the final term leads to difficulties, not the least of
which is the identification of a momentum density that does not satisfy the Heaviside--Larmor symmetry
\cite{Shevchenko2010,Shevchenko2011}\footnote{It is by no means straightforward to obtain the Einstein--Laub
force density, in particular the final hidden--momentm related term, from the microscopic Lorentz force law and 
it has been suggested, for this reason, that the latter is incorrect \cite{Mansuripur2012}.  A relativistic treatment, 
however, reveals that the required hidden--momentum contribution arises quite naturally from the Lorentz force 
law \cite{Vanzella2013,Barnett2013,Saldanha2013,Khorrami2013,Griffiths2013}.}.  The above derivation of the 
Einstein--Laub force density shows how the magnetic terms follow from the polarisation terms by simple symmetry
arguments and so provides a new perspective on the complete force density.  It is interesting to note, moreover, 
that the force density appropriate for a dielectric medium (\ref{LForce}) may be obtained by consideration of the 
action on the individual dipoles making up the medium \cite{BarnettLoudon2006} and that the most direct way
to arrive at the Einstein--Laub force density is to obtain the magnetisation part by treating a collection of Gilbertian
magnetic dipoles \cite{Gilbert1600}.

It is also shown in the original paper \cite{EinsteinLaub1908} that the classical form of the Abraham momentum
and the classical force density satisfy the conservation condition
\begin{equation}
\label{Eq5.5}
\frac{\partial}{\partial t}{\bf G}^{\rm A}(t) = - \int d{\bf r} \: {\bf f}^{\rm EL}({\bf r},t) \, .
\end{equation}
The integration is taken over all space and the relation is valid for fields that vanish at infinity.  This equality of
the rate of change of the Abraham momentum of the light to minus the total Einstein--Laub force on the medium,
or rate of change of material momentum, is as expected on physical grounds and further underlines the identification
of the Abraham momentum with the kinetic momentum of the light \cite{BarnettLoudon2010,Barnett2010}.
Equation (\ref{Eq5.5}) is simply an expression of the conservation of total momentum.

\subsection{Momentum transfer to a half--space magneto--dielectric}

We consider a single--photon pulse normally incident from free space at $z<0$ on the flat surface of a semi--infinite 
magneto--delectric that fills the half space $z>0$.  The pulse is assumed to have a narrow range of frequencies centred 
on $\omega_0$.  Its amplitude and reflection coefficients, the same as in classical theory, are \cite{Jackson1999}
\begin{eqnarray}
R &=& - \frac{\sqrt{\varepsilon/\mu}-1}{\sqrt{\varepsilon/\mu}+1}  \nonumber \\
T &=& \frac{2}{\sqrt{\varepsilon/\mu}+1} \, ,
\end{eqnarray}
where all of the optical parameters depend on the frequency.  The damping parameters in the imaginary parts of 
$\varepsilon$ and $\mu$, as they occur in $R$ and $T$, are assumed negligible but they should be sufficient for the
attenuation length 
\begin{equation}
\ell(\omega)= \frac{c}{2\omega\:{\rm Im}(\sqrt{\varepsilon\mu})}
\end{equation}
to give complete absorption of the pulse over its semi--infinite propagation distance in the medium.

The momentum transfer to the medium as a whole is entirely determined by the free-space single-photon
momenta, before and after reflection of the pulse, as
\begin{equation}
\label{Eq5.8}
\left(1 + R^2\right)\frac{\hbar\omega_0}{c} = \left(\sqrt{\frac{\varepsilon}{\mu}}T^2\right)
\frac{\varepsilon + \mu}{2\eta_{\rm p}}\frac{\hbar\omega_0}{c} \, ,
\end{equation}
with the small imaginary parts of the optical parameters again ignored.  It is often useful to re--express
the momentum transfer in terms of a single transmitted photon with energy $\hbar\omega_0$ at
$z = 0^+$.  This quantity is obtained by removal of the transmitted fraction of the pulse energy,
given by the bracketed term on the right of (\ref{Eq5.8}), as
\begin{equation}
\label{ptotal}
p_{\rm total} = \frac{\varepsilon + \mu}{2\eta_{\rm p}}\frac{\hbar\omega_0}{c}
= \frac{1}{2}\left(\sqrt{\frac{\varepsilon}{\mu}}+\sqrt{\frac{\mu}{\varepsilon}}\right) \, ,
\end{equation}
in agreement with previous work \cite{Mansuripur2007,Shockley1968}.  The remainder of the section
considers the separation of this total transfer of momentum to the magneto--dielectric into its 
surface and bulk contributions.

Previous calculations \cite{LBB2005,Loudon2002} of the radiation pressure on a semi--infinite
dielectric were made by evaluation of the Lorentz force on the material.  This method is generalised
here for a magneto--dielectric medium.  For a transverse plane--wave pulse propagated parallel
to the $z$--axis, with electric and magnetic fields parallel to the $x$ and $y$ axes respectively, the
relevant component of the Einstein--Laub force from (\ref{ELForce}) has the quantised form
\begin{eqnarray}
\label{ELForce1D}
\hat{f}_z^{\rm EL} &=& \mu_0\frac{\partial \hat{P}}{\partial t}\hat{H} -
\frac{1}{c^2}\frac{\partial \hat{M}}{\partial t}\hat{E}  \nonumber \\
&=& \frac{1}{c^2}\left[(\varepsilon - 1)\frac{\partial\hat{E}}{\partial t}\hat{H}
+ \hat{E}(\mu - 1)\frac{\partial\hat{H}}{\partial t}\right] \, .
\end{eqnarray}
The 1D field operators from section \ref{Sect3.3}, given in (\ref{1DOps}), are appropriate here.  
The single--photon pulse is represented by the state
\begin{equation}
|1\rangle = \int d\omega \: \xi(\omega) \hat{a}^\dagger(\omega)|0\rangle \, ,
\end{equation}
where $|0\rangle$ is the vacuum state.  This single--photon state is normalised if we require 
our function $\xi(\omega)$ to satisfy
\begin{equation}
\int d\omega \: |\xi(\omega)|^2 = 1 \, .
\end{equation}
The states, from their construction, satisfy 
\begin{equation}
\hat{a}(\omega)|1\rangle = \xi(\omega)|0\rangle \, .
\end{equation}
A simple choice for the pulse amplitude is
\begin{equation}
\xi(\omega) = \left(\frac{L^2}{2\pi c^2}\right)^{1/4}\exp\left[-\frac{L^2(\omega - \omega_0)^2}{4c^2}\right] 
\end{equation}
with $c/L \ll \omega_0$.  The narrowness of the spectrum of this pulse means that $\omega$ can often
be set equal to $\omega_0$.

The radiation pressure of the magneto--dielectric is obtained by evaluation of the expectation value of
the Einstein--Laub force (\ref{ELForce1D}) for the single--photon pulse.  The normal--order part of the
force operator, indicated by colons, is used to eliminate unwanted vacuum contributions.  A calculation
similar to that carried out in \cite{LBB2005} for a dielectric medium gives
\begin{eqnarray}
\langle 1|:\hat{f}_z^{\rm EL}(z,t):|1\rangle &\approx & 
\left(\frac{\hbar\omega_0}{\sqrt{2\pi}\eta_{\rm p}AL}\right)
\left\{\frac{\varepsilon+\mu}{\ell} \right. \nonumber \\
& & \qquad \qquad \left. -\left[\eta_{\rm p}(\varepsilon+\mu) - 2\eta_{\rm p}\right]
\frac{4c}{L^2}\left(t - \eta_{\rm g}\frac{z}{c}\right)\right\} \nonumber \\
& & \qquad \qquad \times\exp\left[-\frac{2c^2}{L^2}\left(t - \eta_{\rm g}\frac{z}{c}\right)^2
- \frac{z}{\ell}\right] \, ,
\end{eqnarray}
where the real $\varepsilon$, $\mu$, and their derivatives in the group velocity are evaluated at frequency
$\omega_0$ and their relatively small imaginary parts survive only in the attenuation length $\ell$.  It is 
readily verified by integration over $t$ then $z$ that this expression regenerates the total momentum 
transfer in equation (\ref{ptotal}).  The force on the entire material at time $t$ is
\begin{eqnarray}
\langle 1|:\hat{F}_z^{\rm EL}(t):|1\rangle &=& 
A\int_0^\infty dz\langle 1|:\hat{f}_z^{\rm EL}(z,t):|1\rangle 
\approx  \left(\frac{\hbar\omega_0}{\sqrt{2\pi}\eta_{\rm p}\eta_{\rm g}}\right) \nonumber \\
& & \quad \times
\left\{\frac{\eta_{\rm p}}{\eta_{\rm g}\ell}\sqrt{\frac{\pi}{2}}
{\rm erfc}\left[\sqrt{2}\left(-\frac{ct}{L} + \frac{L}{4\eta_{\rm g}\ell}\right)\right] 
\exp\left(-\frac{ct}{\eta_{\rm g}\ell}\right)  \right. \nonumber \\
& & \qquad \left. + \frac{\eta_{\rm g}(\varepsilon+\mu) - 2\eta_{\rm p}}{L}
\exp\left(-\frac{2c^2t^2}{L^2} - \frac{L^2}{8\eta^2_{\rm g}\ell^2}\right) \right\} \, ,
\end{eqnarray}
with appropriate approximations neglecting small terms in the exponents for the long attenuation--length
regime with $L \ll \ell$.  This expression, with error function and exponential contributions, has the same
overall structure as found in other radiation pressure problems \cite{LBB2005,Loudon2002}.  The two terms
in the large bracket are respectively the bulk and surface contributions, with time--integrated values
\begin{eqnarray}
\int_{-\infty}^\infty dt \langle 1|:\hat{F}_z^{\rm EL}(t):|1\rangle &=& p_{\rm total}  \nonumber \\
&=& \underbrace{\frac{\hbar\omega_0}{c}\frac{1}{\eta_{\rm g}}}_{\rm bulk} +
\underbrace{\frac{\hbar\omega_0}{c}\left(\frac{\varepsilon + \mu}{2\eta_{\rm p}}
- \frac{1}{\eta_{\rm g}}\right)}_{\rm surface} \, .
\end{eqnarray}
This is again in agreement with equation (\ref{ptotal}) and it also reduces to equation (5.21) of \cite{LBB2005}
for a non-magnetic material, where the momentum transfer can be written entirely in terms of the phase
and group indices.  The simple Abraham photon momentum again represents that available for transfer to the
bulk material, once the transmitted part of the pulse has cleared the surface, while the more complicated 
surface momentum transfer depends on both $\varepsilon$ and $\mu$, together with their functional forms
embodied in the phase and group refractive indices.  

We conclude this section by noting that the Heaviside-Larmor symmetry retains a presence in all of the
forces and force densities obtained here, in that their forms are unchanged if we interchange, everywhere,
the relative permittivity and permeability.


\section{Conclusion}

Much of the content of the paper presents the generalisations to magneto--dielectrics of results 
previously established for non--magnetic materials with $\mu = 1$.  We believe that the classical
linear response theory in section 2 is novel; it allows, in particular, direct calculation of the electric and 
magnetic field--fluctuation spectra.  The elementary excitations for a medium with multiple electric 
and magnetic resonances are the polaritons, whose phase and group velocities obey generalised
sum rules for magneto--dielectrics \cite{BarnettLoudon2012}.

The quantum theory in section 3 introduces electromagnetic field operators based on the multiple--branch
polariton creation and destruction operators.  It is shown that the vacuum fluctuations of the 
quantised electric and magnetic fields reproduce the spectra obtained from our classical linear--response
theory.  The generalised field operators are shown to satisfy the same required canonical commutation relations
as their simpler counterparts that hold in vacuum.  

The Minkowski and Abraham electromagnetic momentum operators are introduced in section 4 and their
associated single--photon momenta are identified.  The commutators of these momentum and the
vector--potential operators, previously calculated, rely on the canonical commutation relations and, 
through these, on the polariton sum rules.  An extension to angular momentum, both canonical and
kinetic, is achieved by introducing angular momentum densities that are the cross product of the 
position and the Minkowski and Abraham momentum densities respectively.

Throughout our work we are guided by the Heaviside--Larmor symmetry between electric and magnetic
fields.  We show, in section 5, that application of this symmetry leads directly to the Einstein--Laub
force density \cite{EinsteinLaub1908}.  Our final result identifies the surface and bulk contributions in
the force on a semi--infinite magneto--dielectric for the transmission of a single--photon pulse
through its surface.


\ack
This work was supported by the Engineering and Physical Sciences Research Council (EPSRC)
under grant number EP/I012451/1, by the Royal Society and the by Wolfson Foundation.


\end{document}